\documentclass[a4paper]{spie}  
\usepackage[]{graphicx}
\title{Impact of Sodium Layer variations on the performance of the E-ELT MCAO module} 

\author{Schreiber L.\supit{a}, Diolaiti E.\supit{a}, Arcidiacono C.\supit{a}, Pfrommer T.\supit{b}, Holzl{\"o}hner R.\supit{b}, Lombini M.\supit{a} and Hickson P.\supit{c}
\skiplinehalf
\supit{a}INAF - Osservatorio Astronomico di Bologna, Via Ranzani 1, I-40127 Bologna, Italy; \\
\supit{b}European Southern Observatory, Karl-Schwarzschild-Str. 2, D-85748 Garching b. Muenchen, Germany; \\
\supit{c}University of British Columbia, Dept. of Physics and Astronomy, 6224 Agricultural Road, V6T1Z1 Vancouver (BC), Canada
}

\authorinfo{Further author information: (Send correspondence to Laura Schreiber)\\Laura Schreiber: E-mail: laura.schreiber@oabo.inaf.it, Telephone: +39 051 209 5704}

\begin{document} 
  \maketitle 

\begin{abstract}
Multi-Conjugate Adaptive Optics systems based on sodium Laser Guide Stars may exploit Natural Guide Stars to solve intrinsic limitations of artificial beacons (tip-tilt indetermination and anisoplanatism)  and to mitigate the impact of the sodium layer structure and variability. The sodium layer may also have transverse structures leading to differential effects among Laser Guide Stars. 
Starting from the analysis of the input perturbations related to the Sodium Layer variability, modeled directly on measured sodium layer profiles, we analyze, through  a simplified end-to-end simulation code, the impact of the low/medium orders induced on global performance of the European Extremely Large Telescope Multi-Conjugate Adaptive Optics module MAORY. 
\end{abstract}



\section{INTRODUCTION}
\label{sec:intro}  
The future giant telescopes, such as the European Extremely Large Telescope (E-ELT\cite{E-ELT}), the Thirty Meter Telescope (TMT\cite{TMT}) or the Giant Magellan Telescope (GMT\cite{GMT}), will exploit multiple Laser Guide Stars (LGS) as reference sources for the Adaptive Optics (AO) systems in order to ensure an high sky coverage. 
On the other hand, AO systems based on LGS may need Natural Guide Stars (NGS) to solve intrinsic limitations of artificial beacons, such as tip-tilt indetermination and, in the case of Multi-Conjugate Adaptive Optics (MCAO), tip-tilt anisoplanatism\cite{rigaut1992}. 
Moreover recent measurements\cite{tommaso2010} of the sodium layer show a structured density profile and a dynamical evolution even on small spatial and temporal scales\cite{tommaso2014}. The mean sodium altitude variation can be of the order of hundreds of meters or even kilometers. 
On a large diameter telescope this variation has an important effect on the wavefront error (WFE). For example on a 40 m telescope every meter of change of the sodium layer mean altitude translates into $\sim$ 7 nm of defocus\cite{herriot} (RMS WFE).
According to the above discussion, a fast NGS WaveFront Sensor (WFS) measuring Tip-Tilt and Focus is required. Interesting techniques have been
proposed to mix the focus measurements provided by LGS and NGS\cite{herriot}.
As already mentioned, the sodium layer may also have transverse structures across scales
comparable to the linear separation among the artificial beacons in a typical multi-LGS constellation. Depending on the
significance of these transverse structures, more than one NGS could be required to provide a focus reference in an
extended field of view (FoV) AO system.

An additional interesting effect arises from the combination of the finite LGS WFS FoV and of the sodium sodium layer structure and variability.  
If on 8-meter telescopes the main effect is fast focus variation, on a 40-meter class telescope, as a consequence of the truncation of the LGS image due to the sodium layer perspective elongation in combination with the finite FoV of the LGS WFS, additional spurious aberrations are generated\cite{diolaiti2012}. Secondary peaks of non-gaussian (and in general non-symmetric) sodium profiles could be 'truncated' in the sub-apertures far from the laser launcher, where the spots are more elongated, producing a kind of discontinuity in the slope measurements across the pupil. This translates into more complex wavefront than pure defocus and, in the case of edge projection, tip-tilt.
These aberrations are only due to the sodium layer and have no relation with the wavefront aberrations due to atmospheric turbulence that the AO system should compensate. The variability of the sodium profile on short time scales prevents the possibility to take into account for this effect as a simple slope offset calibration measuring the averaged centroids over an adequate time interval to filter out the residual atmospheric turbulence.
A Reference WFS, independent from sodium layer issues and based on NGS, is therefore maybe required to monitor these spurious aberrations, especially if a large spot truncation is foreseen. 

All these aspects show up in MAORY\cite{maory}, a concept for a MCAO module for the E-ELT designed to provide uniform correction over an extended FoV using 6 Sodium LGS for wavefront sensing. 

Starting from the analysis and modeling of the input perturbations related to the Sodium Layer variability (Sect. \ref{sodium}), we analyze the impact of the low/medium orders induced on global performance of the E-ELT MCAO module. We designed for this purpose a simplified end-to-end simulation code (Sect. \ref{E2Ecode}) able to inject in the system time series of low order spurious aberrations related to the density profile variation of each LGS. Due to the simplicity of the code, time series of sodium profiles of the order of few hours can be analyzed in few minutes. A full end-to-end simulation code, able to simulate all the relevant aspects of the system, would take much more time to simulate such a long time sequence.  
The analyzed time series have been generated synthetically from the analysis of the temporal Power spectra of each spurious low order mode using as starting point directly measured sodium layer profiles, taken with a high resolution Lidar system\cite{tommaso2010} on the 6-meter Large Zenith Telescope (LZT) near Vancouver. For this paper we did not consider the spatial structure variations among different LGSs, leaving this upgrade for a future work. We simply replicated the same sodium profile sequence for all the 6 LGSs and considering three NGSs for the Reference WFS measurements. 
This 'exercise' gives us anyway the possibility to have an idea on the impact of the sodium profile variation in terms of WFE RMS on the MAORY error budget depending on the design of the Reference WFS in terms of number of sub-apertures and loop frequency. 

\section{SODIUM DATA REDUCTION} \label{sodium}

   \begin{figure}
   \begin{center}
   \begin{tabular}{c}
   \includegraphics[height=7cm]{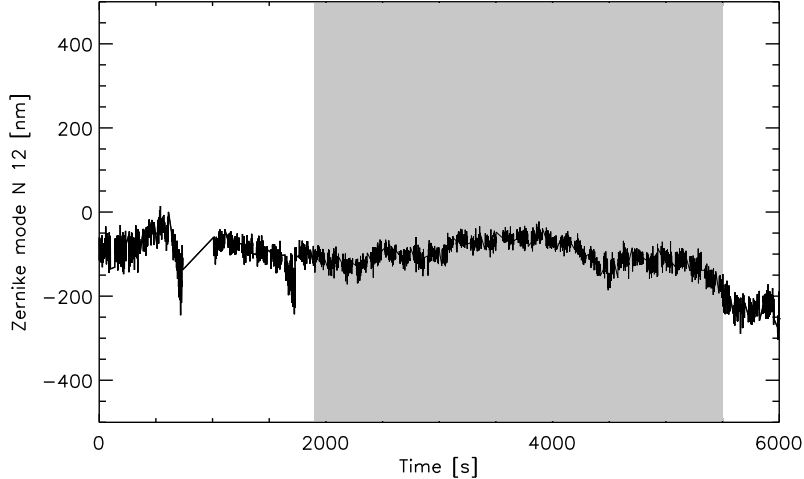}
   \end{tabular}
   \end{center}
   \caption[coeff_seq] 
   { \label{fig:coeff_seq} 
Sample time series of Zernike mode amplitude. The gray region highlights the selected data portion of 1 hour for that night.}
   \end{figure}

   \begin{figure}
   \begin{center}
   \begin{tabular}{c}
   \includegraphics[height=7cm]{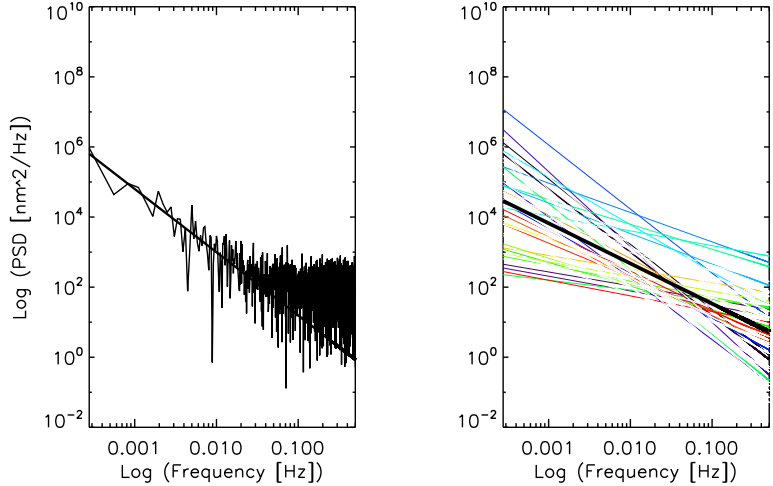}
   \end{tabular}
   \end{center}
   \caption[psd_fit] 
   { \label{fig:psd_fit} 
\textit{Left panel}: Example of a power spectrum extracted from the 1 hour portion of the sequence of Fig. \ref{fig:coeff_seq} highlighted in grey. \textit{Right panel}: All power the spectra relative to the Zernike mode 12, spanning the dynamical
range in the power-frequency plot.}
   \end{figure} 

We analyzed a set of sodium profile measurements\cite{tommaso2010} including more than 20 nights spanned on 2 years between 2008 and 2009. The original data were provided with a vertical resolution of 4 meters and a temporal resolution of 20 ms in order to trace the sodium layer dynamics on time scales important to AO systems. The data were divided in packages of few minutes divided by few seconds gaps required for data transfer. In order to reduce the impact of photon noise the data were binned in time to final temporal sampling of 1 s and in vertical resolution to final spacial sampling of few tens of meters.
In a previous publication\cite{diolaiti2012} we described how we used the sodium data to compute the wavefront aberrations measured by an ideal LGS WFS with $80 \times 80$ sub-apertures and a given FoV (10'', 15'' and 20''). In a few words, we computed the angular centroid of the sodium profile as seen from each of the WFS sub-apertures with and without FoV truncation. The translation of this differential angular centroid map in X-Y slopes on the pupil plane was then converted into about a hundred of Zernike polynomials coefficients by modal fitting. For more details on this procedure, please refer to the cited previous paper. 

We therefore obtained for each night, long time series of Zernike coefficients characterized by larger amplitudes for more severe truncation. The Zernike excited orders depends on the laser launching scheme and on the position of the LGS respect to the entrance pupil. The amplitude of each coefficient can be of the order of hundreds of nanometers, becoming negligible for high order modes. Fig. \ref{fig:coeff_seq} depicts a sample series of the Zernike mode number 12 (excluding piston and being tilt numbered as 0) obtained from one night measurements. We filled the data transfer gaps interpolating linearly between the last value of the previous package and the first of the next package.  

Since we were interested in higher temporal frequencies, we decided to move from a temporal domain to a frequency domain analysis. 
A frequency domain approach allowed us to extrapolate the power spectra beyond the noise floor and thus avoid underestimating the impact of high temporal frequencies. 
Before computing the power spectra, the times series were split into about 60 packages of 1 hour each, excluding very noisy parts (the data included also measurements taken when the whether conditions were not ideal for astronomical observations). We also avoided the spikes due to meteor trails, to prevent the spikes themselves from contaminating the power spectra. The gray region of Fig. \ref{fig:coeff_seq} highlights the selected data portion of 1 hour for that night. Power spectra were then computed for each data package and for each Zernike mode and were extrapolated to high frequencies by fitting with a power low. 
The power spectrum represented in Fig. \ref{fig:psd_fit} has been computed using the time sequence in the gray region of Fig. \ref{fig:coeff_seq}. All the power spectra relative to all the 1 hour packages spanning the dynamical range in the power-frequency plot concerning the 12th Zernike mode are grouped in the right panel of Fig. \ref{fig:psd_fit}. The thickest line represents the power low averaged over the 60 data packages. 

The average power spectrum of each mode can be used to generate random sequences of Zernike coefficients at the desired frequency. These coefficients can be combined to build a sequence of spurious aberrations generated by a single LGS. 


\section{THE SIMPLIFIED SIMULATION TOOL} \label{E2Ecode}
\subsection{Concept}
   \begin{figure}
   \begin{center}
   \begin{tabular}{c}
   \includegraphics[height=7cm]{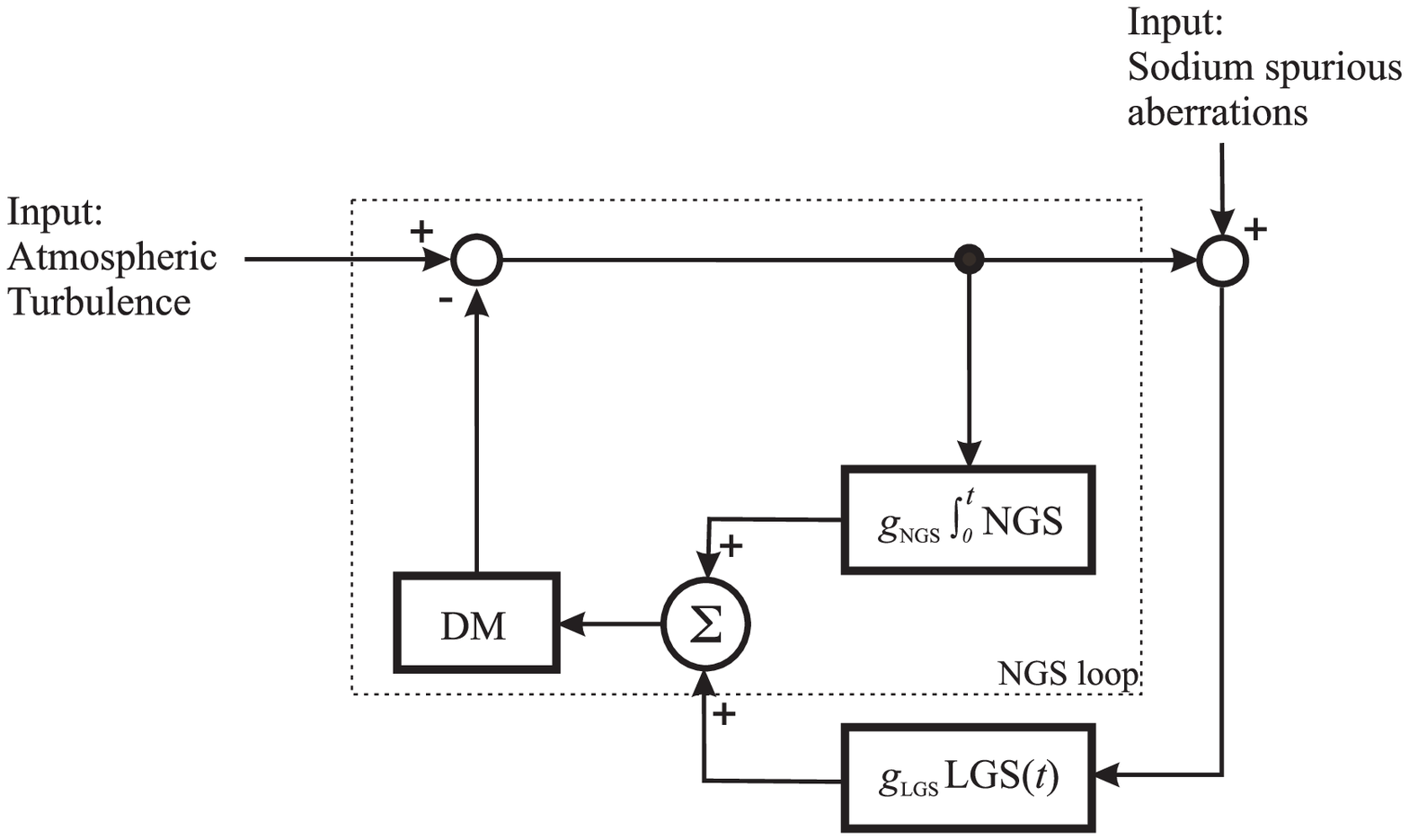}
   \end{tabular}
   \end{center}
   \caption[schema concettuale] 
   { \label{fig:schema concettuale} 
Simulation code conceptual scheme.}
   \end{figure} 
	
In order to evaluate the impact of the low/medium orders induced by sodium layer variations on the MAORY performance, we designed a fast and simplified numerical simulation tool. It has been developed in the Interactive Data Language (IDL) version 8.2.3.
Due to its simplicity, this code is able to simulate long time intervals (tens of minutes) in a short time (tens of seconds). 
The code propagates the spurious aberrations introduced in single-LGS wavefronts by the spot truncation (see Sect. \ref{sodium}) in a layer-oriented fashion by using simple geometrical approximations and it combines them in a numerical way. 
Atmospheric turbulence is not considered: the code is designed to study the propagation of the spurious aberrations induced by the LGS WFS and how these aberrations can be controlled by a NGS-based WFS (so called Reference WFS), with relatively slow temporal rate and coarse spatial sampling.
The code is based on two AO loops, one (faster) for the LGS WFS and one (slower) for the Reference WFS. The first 5 Zernike modes (tip-tilt, defocus and astigmatism\footnote{astigmatism is assumed to be measured by the NGS TTF WFS: in principle a WFS measuring focus also provides the necessary information to retrieve astigmatism, at the expense of a slightly larger WFE due to noise propagation}) are set to 0 in the LGS signals, therefore these modes are not propagated into the AO loop. These modes are assumed to be measured perfectly by a fast tip-tilt Focus Astigmatism (TTFA) NGS WFS, which is not included in the simulation for simplicity.
The performance of the system is expressed in terms of WFE evolution computed for each LGS WFS time evolution step.
For every temporal step of the \textit{main} loop (set by the LGS WFS integration time) the code computes the injected spurious wavefronts at the DMs conjugation altitudes by shifting and averaging the single-channel signals in a layer-oriented fashion taking into account the cone effect and subtracting the previously computed DM correction. For every temporal step of the \textit{auxiliary} loop (set by the NGS Reference WFS integration time) the code computes the wavefront measured by each NGS by coadding the portions of the DMs illuminated by the NGS itself. Low-medium order wavefront reconstruction on the DM planes is then performed by combining the NGS wavefronts at the conjugation altitudes following a layer-oriented approach. The LGS and the NGS measurements are then combined together and used to compute the correction. In other words, the NGS WFS continuously updates a reference slope offset, to be subtracted from the LGS measurements before sending the commands to the DMs. 
Fig. \ref{fig:schema concettuale} shows a diagram of the simulation tool.      

\subsection{Simulation Inputs}
The simulation code accepts 2 main inputs:
\begin{itemize} 
\item An IDL data structure containing the telescope and the AO system main parameters, (such as diameter, central obstruction, number of DMs, conjugation altitudes, LGS launching angles and positions, NGS direction) and the parameters to be optimized (NGS WFS integration time and NGS WFS number of sub-apertures). In order to fasten the code, the structure contains also all the arrays and constants (such as masks, interaction matrices, parameters concerning the geometry of the problem) that do not need to be computed in closed loop. 
\item A sequence of spurious aberrations generated by the sodium layer computed as described in Sect. \ref{sodium}. In order to lighten the overall computation it is preferable to previously compute the 6 LGSs combined wavefront distortion. A separated package is available for this purpose. 
\end{itemize}

\subsection{Results} 

   \begin{figure}
   \begin{center}
   \begin{tabular}{c}
   \includegraphics[height=7cm]{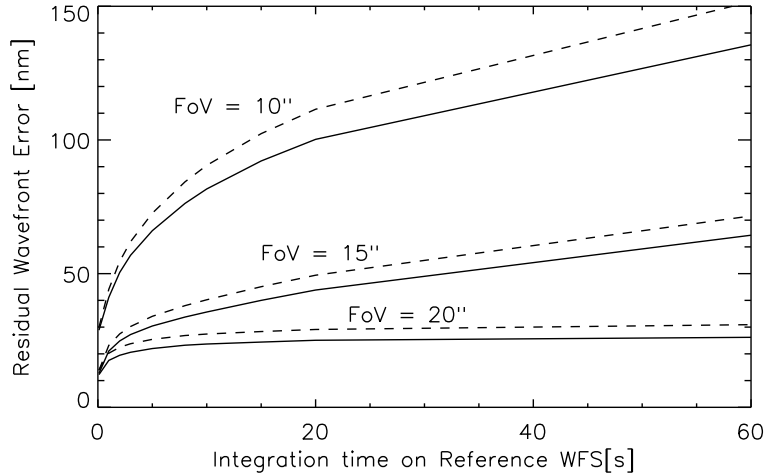}
   \end{tabular}
   \end{center}
   \caption[wfe_avg] 
   { \label{fig:wfe_avg} 
Residual WFE as a function of the Reference WFS sampling time, assuming three different values of
LGS WFS FoV and assuming that the Reference WFS is a $10 \times 10$ Shack-Hartmann WFS. Continuous lines refer to the WFE measured at the MAORY FoV center. Dashed lines refer to the average WFE over four points at the four corners of the MICADO FoV.}
   \end{figure} 
   \begin{figure}
   \begin{center}
   \begin{tabular}{c}
   \includegraphics[height=7cm]{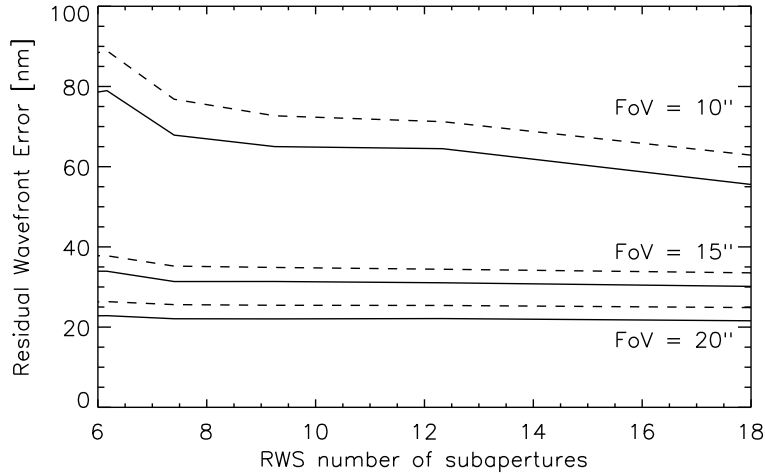}\\
   \end{tabular}
   \end{center}
   \caption[wfe_nsub] 
   { \label{fig:wfe_nsub} 
Residual WFE as a function of the Reference WFS number of \textit{n} sub-apertures (where \textit{n} refers to the side of the SH WFS) assuming an integration time of 5 s. Continuous lines refer to the WFE measured at the MAORY FoV center. Dashed lines refer to the average WFE over four points at the four corners of the MICADO FoV.}
   \end{figure} 
	
	\begin{table}
\caption{Main parameters of the simulation.}  
\label{table:simulation} 
\centering 
\begin{tabular}{||c c ||} 
\hline 
Telescope diameter &  37 m \\ 
Sub-aperture diameter & 0.5 m  \\
Number of LGS &  6   \\
LGS launching angle (radius) & 60" \\
LGS dist from telescope edge & 1 m \\
Number of NGS &  3   \\
NGS direction (radius) &   80"   \\
(equilateral triangle asterism) & \\
DM conjugation altitudes &  0, 4000, 12700 m   \\
\hline 
\end{tabular}
\end{table}

The main goal of this code is to compute the WFE due to the combination of the sodium spurious aberrations seen by each single LGS WFS and to evaluate the performance in closed loop of a slow Reference WFS with few sub-apertures. The NGS loop integration time and the number of required sub-apertures are key parameters strongly related to the sky coverage.
As in the previous publication\cite{diolaiti2012}, we run the simulation for three values of LGS WFS FoV: 10", 15", 20".  
For each case the analysis was repeated for different numbers of NGS Reference WFS sub-apertures (starting from $6 \times 6$) and for different integration times ranging from 0.1 s to 60 s.
We summarized in Table \ref{table:simulation} the main input parameters common to all the analyzed cases. The performance have been evaluated in 5 positions across the MAORY FoV: one in the center and four at the four corners of the MICADO\cite{davies2010} FoV ($\sim$ 37" far from the FoV center). We considered the NGSs positioned at the vertexes of an equilateral triangle at the edge of the research field, but their direction can be arbitrary changed.  

The results are shown in Fig. \ref{fig:wfe_avg} for a $10 \times 10$ NGS Reference WFS. As already found in previous publication, the results show a relatively strong dependency on the LGS WFS FoV: while 15" and 20" FoV lead to reasonable residual WFE, the case of 10" seems again to induce large errors for integration times of few seconds, therefore which could affect sky coverage. This error seems not to be mitigated by the tomographic reconstruction process. A possible explanation is that, since the positions of the laser launchers are different respect to the entrance pupil, the elongation pattern of each LGS WFS is rotated respect to the others causing the sodium induced aberrations to be different for each LGS, even when there is no horizontal variation in the sodium density profile. However the use of noise priors in the tomographic reconstruction is expected to attenuate\cite{miska2013} the impact of measurements from sub-apertures with large spot elongation and therefore severe truncation effects. A comforting result is depicted in Fig. \ref{fig:wfe_nsub}. The Reference WFS integration time is fixed to 5 seconds and the number of sub-apertures varies starting from $6 \times 6$ to larger numbers. This plot tells that the majority of the WFE is due to low order aberrations, well measurable with a $7 \times 7$ WFS as well as a $10 \times 10$ one.

\section{CONCLUSIONS}

We have presented a simplified simulation code written in IDL able to evaluate the impact of the low/medium orders induced by sodium layer variations on the performance of the future E-ELT MCAO module MAORY. 
A Reference WFS, independent from sodium layer issues and based on NGSs, is maybe required to monitor these spurious aberrations, especially if a large spot truncation is foreseen. 
The main scope of this work was to describe the code and to present its potentialities. As a first exercise we have run the code on a set of sodium profile measurements including more than 20 nights spanned on 2 years between 2008 and 2009. A lot of new sodium data are already available and we will soon run our code on a larger sample. 
Moreover we replicated the same sodium profile sequence for all the 6 LGSs, thus not considering the horizontal sodium layer variations. The code is anyway able to work with a different profile for each LGS.
We used the sodium data to compute the wavefront aberrations measured by an ideal 74 $\times$ 74 LGS WFS and a given FoV (10", 15" and 20"), that represent, together with the main parameters of the system, the starting point of our code. 
Through this kind of analysis it is possible to trace an high-level design of a Reference NGS WFS, in terms of required number of sub-apertures and integration time. From the obtained results it is apparent a relatively strong dependency on the LGS WFS FoV, especially in the most truncated case considered in this work. 
In order to compensate for the large amount of WFE introduced by the low/medium orders, in case of a small sub-aperture FoV the Reference WFS integration time has to be as small as possible, thus compromising the high sky coverage requirement. On the other hand, a smaller number of sub-apertures seems to work good as well, partially compensating the necessity to reduce the integration time. The analyzed sample indicates that in case of sub-aperture FoV $\geq$ 20", the WFE due to spot truncation is acceptable and does not need to be compensated. However this represents a huge FoV, requiring a very large number of pixels in the LGS WFS detector. 
We anyway remember that the use of noise priors in the tomographic reconstruction is expected to strongly attenuate\cite{miska2013} the impact of measurements from sub-apertures with large spot elongation and therefore severe truncation effects. We are therefore planning to add in the code the possibility to supply also noise priors.

\acknowledgments     
This work was partly supported by the “Progetto Premiale E-ELT 2012” (PI Monica Tosi) funded by the Italian Ministry for Education, University and Research.
%
%
\bibliography{spie2014}   

\begin{thebibliography}{10}

\bibitem{E-ELT}
R.~{Gilmozzi} and J.~{Spyromilio}, ``{The 42m European ELT: status},'' in {\em
  Society of Photo-Optical Instrumentation Engineers (SPIE) Conference Series},
   {\em Society of Photo-Optical Instrumentation Engineers (SPIE) Conference
  Series} {\bf 7012}, Aug. 2008.

\bibitem{TMT}
L.~{Simard}, D.~{Crampton}, B.~{Ellerbroek}, and C.~{Boyer}, ``{The
  instrumentation program for the Thirty Meter Telescope},'' in {\em Society of
  Photo-Optical Instrumentation Engineers (SPIE) Conference Series},  {\em
  Society of Photo-Optical Instrumentation Engineers (SPIE) Conference Series}
  {\bf 8446}, Sept. 2012.

\bibitem{GMT}
M.~{Johns}, P.~{McCarthy}, K.~{Raybould}, A.~{Bouchez}, A.~{Farahani},
  J.~{Filgueira}, G.~{Jacoby}, S.~{Shectman}, and M.~{Sheehan}, ``{Giant
  Magellan Telescope: overview},'' in {\em Society of Photo-Optical
  Instrumentation Engineers (SPIE) Conference Series},  {\em Society of
  Photo-Optical Instrumentation Engineers (SPIE) Conference Series} {\bf 8444},
  Sept. 2012.

\bibitem{rigaut1992}
F.~{Rigaut} and E.~{Gendron}, ``{Laser guide star in adaptive optics - The tilt
  determination problem},'' {\em A\&A}~{\bf 261}, pp.~677--684, Aug. 1992.

\bibitem{tommaso2010}
T.~{Pfrommer} and P.~{Hickson}, ``{High-resolution mesospheric sodium
  observations for extremely large telescopes},'' in {\em Society of
  Photo-Optical Instrumentation Engineers (SPIE) Conference Series},  {\em
  Society of Photo-Optical Instrumentation Engineers (SPIE) Conference Series}
  {\bf 7736}, July 2010.

\bibitem{tommaso2014}
T.~{Pfrommer} and P.~{Hickson}, ``{High resolution mesospheric sodium
  properties for adaptive optics applications},'' {\em A\&A}~{\bf 565},
  p.~A102, May 2014.

\bibitem{herriot}
G.~{Herriot}, P.~{Hickson}, B.~{Ellerbroek}, J.-P. {V{\'e}ran}, C.-Y. {She},
  R.~{Clare}, and D.~{Looze}, ``{Focus errors from tracking sodium layer
  altitude variations with laser guide star adaptive optics for the Thirty
  Meter Telescope},'' in {\em Society of Photo-Optical Instrumentation
  Engineers (SPIE) Conference Series},  {\em Society of Photo-Optical
  Instrumentation Engineers (SPIE) Conference Series} {\bf 6272}, July 2006.

\bibitem{diolaiti2012}
E.~{Diolaiti}, L.~{Schreiber}, I.~{Foppiani}, and M.~{Lombini}, ``{Dual-channel
  multiple natural guide star wavefront sensor for the E-ELT multiconjugate
  adaptive optics module},'' in {\em Society of Photo-Optical Instrumentation
  Engineers (SPIE) Conference Series},  {\em Society of Photo-Optical
  Instrumentation Engineers (SPIE) Conference Series} {\bf 8447}, July 2012.

\bibitem{maory}
E.~{Diolaiti}, ``{MAORY: A Multi-conjugate Adaptive Optics RelaY for the
  E-ELT},'' {\em The Messenger}~{\bf 140}, pp.~28--29, June 2010.

\bibitem{davies2010}
R.~{Davies} and R.~{Genzel}, ``{MICADO: The Multi-adaptive Optics Imaging
  Camera for Deep Observations},'' {\em The Messenger}~{\bf 140}, pp.~32--33,
  June 2010.

\bibitem{miska2013}
M.~{Le Louarn}, C.~{B{\'e}chet}, and M.~{Tallon}, ``{Of spiders and elongated
  spots},'' in {\em Proceedings of the Third AO4ELT Conference},  S.~{Esposito}
  and L.~{Fini}, eds., Dec. 2013.

\end{thebibliography}
\bibliographystyle{spiebib}   

\end{document}